\documentclass[twocolumn,secnumarabic,amssymb, nobibnotes, aps, prd]{revtex4-1}

\usepackage{upgreek} % to have roman greek letter

\usepackage{epsfig, amsmath,amsfonts, amssymb,graphicx,amsthm,color}

\usepackage{mathtools}% For \MoveEqLeft
\usepackage{dcolumn}% Align table columns on decimal point
\usepackage{bm}% bold math
\usepackage{rotating}
\usepackage{amssymb}
\usepackage[latin1]{inputenc}
\usepackage{comment}
\usepackage{multirow}
\usepackage{tabularx}
\usepackage[pagebackref=false]{hyperref}

\hypersetup{
    colorlinks=true,       
    linkcolor=blue,          
    citecolor=blue,       
    filecolor=magenta,      
    urlcolor=blue          
}
\newcommand {\Ket}[1]         {\ensuremath{| \, #1 \rangle}} 
\newcommand {\Bra}[1]         {\ensuremath{\langle #1 \,|}} 
\graphicspath{{Images/}} 
\usepackage{lipsum}
\usepackage[symbol]{footmisc}

\usepackage{filecontents}
{% for the vertical padding

\begin{document}

\title{Isolated core excitation of high orbital quantum number Rydberg states of ytterbium}
\author{Henri Lehec}
\altaffiliation[Present address: ]{Institut f\"ur Physik, Univ. Mainz, Staudingerweg 7, 55128 Mainz, Germany}
\email{hlehec@uni-mainz.de}

\author{Xin Hua}
\altaffiliation[Present address: ]{Institut de Physique de Nice (INPHYNI), Parc Valrose, 06108 Nice, France}
\email{xin.hua@unice.fr}
\author{Pierre Pillet}
\author{Patrick Cheinet}
\email[Contact: ]{patrick.cheinet@u-psud.fr}

\affiliation{Laboratoire Aim\'{e} Cotton, CNRS, Univ. Paris-Sud, Universit\'e Paris-Saclay, B\^{a}t. 505, 91405 Orsay, France }
\date{\today}

\begin{abstract}
\noindent\textbf{Abstract} We study isolated core excitation of ultra cold ytterbium Rydberg atoms of high orbital quantum number. Measurements were performed on the $6s_{1/2} 40l \rightarrow 6p_{1/2} 40l $ transition with $l=5-9$. The extracted energy shifts and autoionization rates are in good agreement with a model based on independant electrons, taking into account interactions in a perturbative approach. We reveal a particularly long persistence of the autoionization rates with the orbital quantum number, explained by the strong coupling of the $6p_{1/2}nl$ autoionizing state with the $5d_{3/2}\epsilon l'$ continua compared to previously studied divalent atoms.
\end{abstract}

\maketitle

\section{Introduction}

Rydberg atoms offer an ideal platform for quantum simulation experiments \cite{WEI10}. Thanks to their strong, long-range and tunable interactions, they are good candidates to study many-body phenomena. Atoms trapped in optical lattices have been used to demonstrate spatial correlations in the Rydberg excitation \cite{SCH12} and more recently, tunable arrays of Rydberg atoms have been used to demonstrate the simulation of Ising spin systems \cite{LAB16,BER17} or topological matter \cite{DEL19}. In these experiments, the imaging technique relies on the fluorescence of the atoms after returning them to their ground state. A drawback of this technique is the false-positive detection of residual non-evacuated ground state atoms. This typically sets a limit on the maximum number of atoms to few hundred, which is detrimental for future quantum simulation experiments. 

We investigate here the possibility to image the Rydberg atoms through the fluorescence induced by isolated core excitation (ICE) \cite{COO78}. Permitted in multivalent atoms such as strontium, barium, ytterbium \textit{etc.}, this technique could allow efficient and accurate in-situ imaging of the Rydberg atoms. However, due to the interaction between the two valence electrons, doubly excited Rydberg atoms may rapidly autoionize \cite{MCQ13,LOC13}. A first method to reduce autoionization is to use high principal quantum number states \cite{COO78,FIE18}. The autoionization is known to evolve as $1/n^3$, following the Rydberg wave function amplitude evolution close to the core. Another method is to use the strong centrifugal barrier of high orbital quantum number states that prevents the Rydberg electron to interact with the ionic core. So far, this technique has been demonstrated in barium \cite{JON88} and in strontium \cite{COO78}. In these two atoms, the autoionization has been found to evolve close to the law $10^{-l}$ but unlike the evolution with $n$, this law is not universal and cannot be demonstrated theoretically.\\

\indent In this article, we study the isolated core excitation on the $6s_{1/2} 40l  \rightarrow 6p_{1/2} 40l $ ionic core (IC) transition of high orbital quantum number $l=5-9$ Rydberg states of $^{174}$Yb in ultra-cold conditions. We show a good agreement between our measurements and a simple theoretical model based on independant electrons, where interactions are treated perturbatively. We also reveal that in ytterbium, the coupling of the $6p_{1/2}nl$ autoionizing state with the $5d_{3/2}\epsilon l'$ continua is particularly strong. Thus the autoionization rate decay with the orbital quantum number $l$ is relatively weak, compared to that of barium or strontium.

\section{ \label{Sec2} Experimental methods}
\begin{figure}[t]
	\includegraphics[scale=0.25]{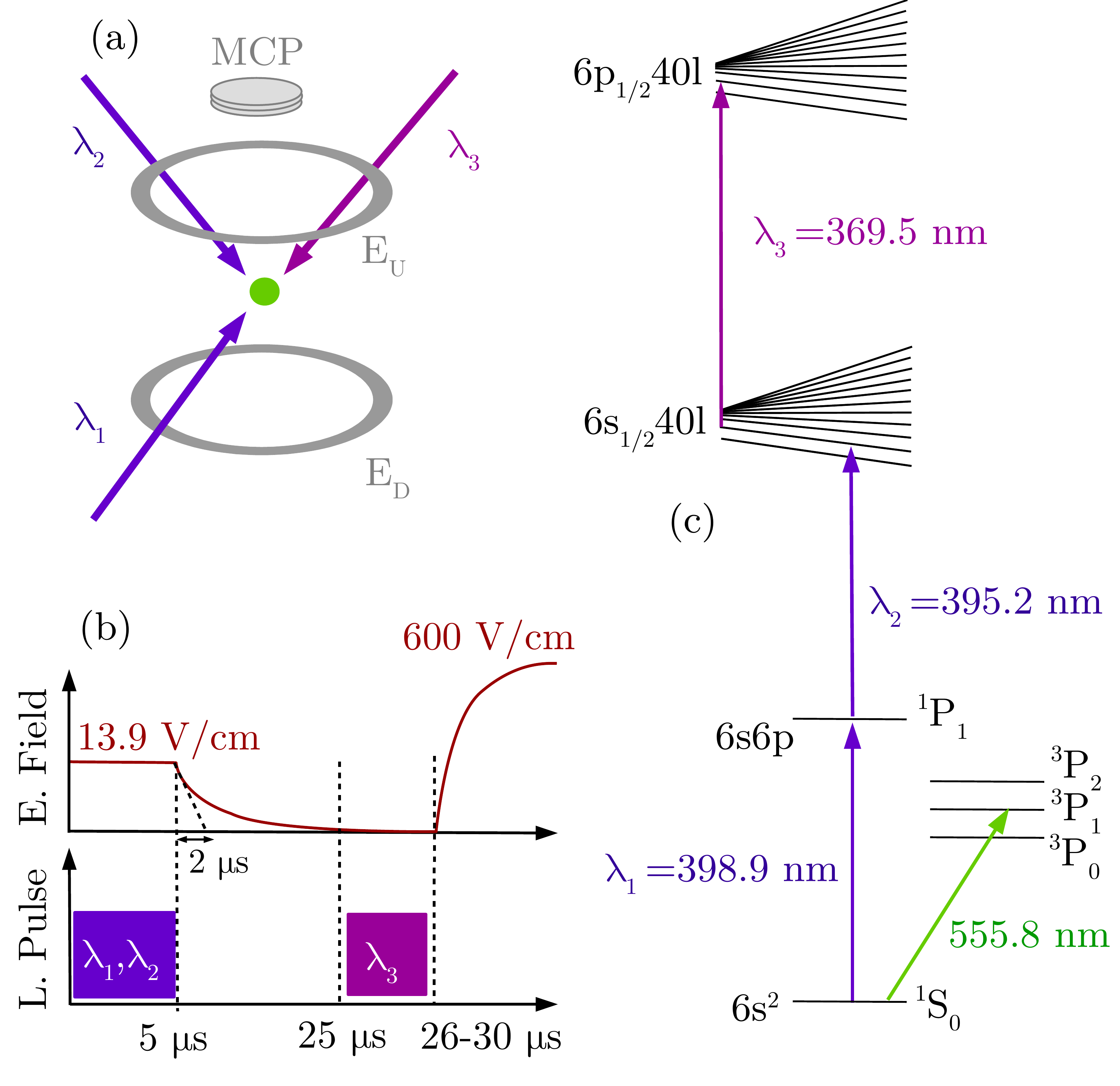}
	\caption{\label{Sch} (a) Schematic of the experimental setup. (b) Time sequence of the experiment. A cold cloud of ytterbium atoms is laser excited ($\lambda_1 + \lambda_2$) to a high-$l$ Rydberg state under a static electric field (electrode $E_U$). After the adiabatic ramp-down of the electric field, a third laser $\lambda_3$ is applied to excite the isolated core transition. Resulting autoionized atoms are guided towards an MCP detector thanks to electrode $E_D$.  (c) Scheme of energy levels involved in this experiment.}
\end{figure}

The schematic of the experiment is represented in \hyperref[Sch]{figure \ref{Sch}}. Cold ${}^{174}$Yb atoms are continuously loaded in a 3D magneto-optical-trap (MOT) operating on the intercombination line $6s^2 \ {}^1 S_0 \rightarrow 6s6p \ {}^3 P_1$ at 555.8 nm. The atomic cloud contains about 1 $\times 10^6$ atoms and has a size of about 1 mm.
In order to selectively excite the $6s nl$ state with high orbital quantum number $l$, we use the so-called \emph{Stark-Switching} technique \cite{FRE76}. The Rydberg excitation is performed with 2 photons through the intermediate $6s6p \ {}^1 P_1$ level. This first level is reached thanks to an external cavity diode laser (ECDL) locked on the $6s^2 \ {}^1 S_0 \rightarrow 6s6p \ {}^1 P_1$ resonance, collimated with a 1 mm waist and with a peak intensity of about 30 mW/cm$^2$ on the atoms. The second photon is provided by a cavity doubled Titanium-Saphire laser at around 395 nm and has a waist of about 0.5 mm and a peak intensity of 50 W/cm$^2$. Both lasers are pulsed on simultaneously during 5 $\mu$s.

\begin{figure}[b]
	\includegraphics[scale=0.32]{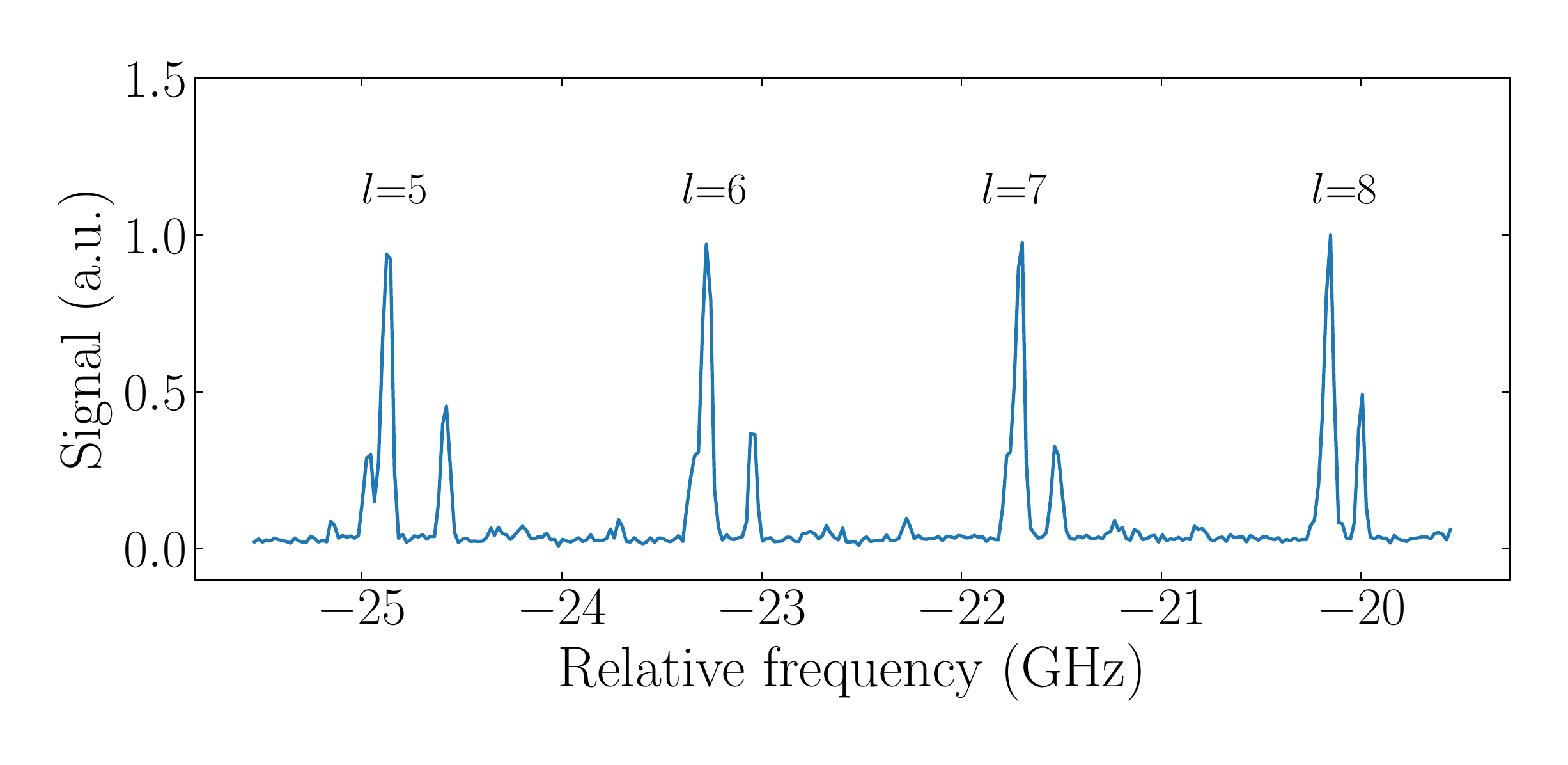}
	\caption{\label{Stark} Rydberg excitation spectrum under a static electric field of 13.9 V/cm of the states connecting to $n=40,l=5$ to $l=8$ levels. Frequencies are given relatively to the center of the $n$=40 multiplicity. For the lowest $l$ states a sub-structure of 3 peaks can be seen. For every $l$ state, the strongest peak was excited.}
\end{figure}

The two-photon excitation is performed under a static electric field in order to authorize the excitation of states connecting to the different $6s40l$ levels, as can be seen in \hyperref[Stark]{figure \ref{Stark}}. The states appear as groups of 2 or 3 peaks and we choose to excite the atoms at the strongest peak for each group. Computation of the exact Stark structure of a divalent atom has not been performed yet, which prevents predicting to which level it connects at zero electric field. Nevertheless, it was demonstrated that they adiabatically connect to states of known orbital momentum \cite{FRE76}. The electric field is then ramped down with a 2 $\mu$s time constant to minimize diabatic transfers towards adjacent $6s40 l\pm 1$ states. The total ramp-down time of 20 $\mu$s is chosen shorter than the Rydberg levels total lifetime. In order to avoid unwanted residual Stark mixing of the $6s40l$ states at the end of the ramp, 8 electrodes are used to compensate stray electric fields at an accuracy below 10 mV/cm. 

Afterwards, isolated core excitation is performed on the $6s_{1/2}40 l \rightarrow 6p_{1/2}40l$ transition at 369.5 nm thanks to an ECDL laser which is pulsed on during a variable duration of 1 - 5 $\mu$s with an AOM. The durations were chosen to produce a significant autoionization signal, without exhausting the Rydberg cloud. The intensity on the atoms is around 10 mW/cm$^2$ and the wavelength is measured thanks to a calibrated wavemeter High-Finesse WSU providing an accuracy of 10 MHz. An electric field ramp is then applied on the $E_D$ electrode in order to guide autoionized atoms towards a micro-channel plate (MCP) detector to be counted as $N_{ai}$. After around 2 $\mu$s, the electric field reaches the ionization threshold of the Rydberg levels. The resulting ions are subsequently guided to the same detector and arrive at a different time, allowing to determine the number of residual Rydberg atoms and thus the total number of atoms initially excited to the $6s_{1/2}40l$ Rydberg state ($N_{tot}$). 

\section{ \label{Sec3} Experimental results and analysis method}
\label{ExpRes}
\begin{figure}[b]
	\includegraphics[scale=0.54]{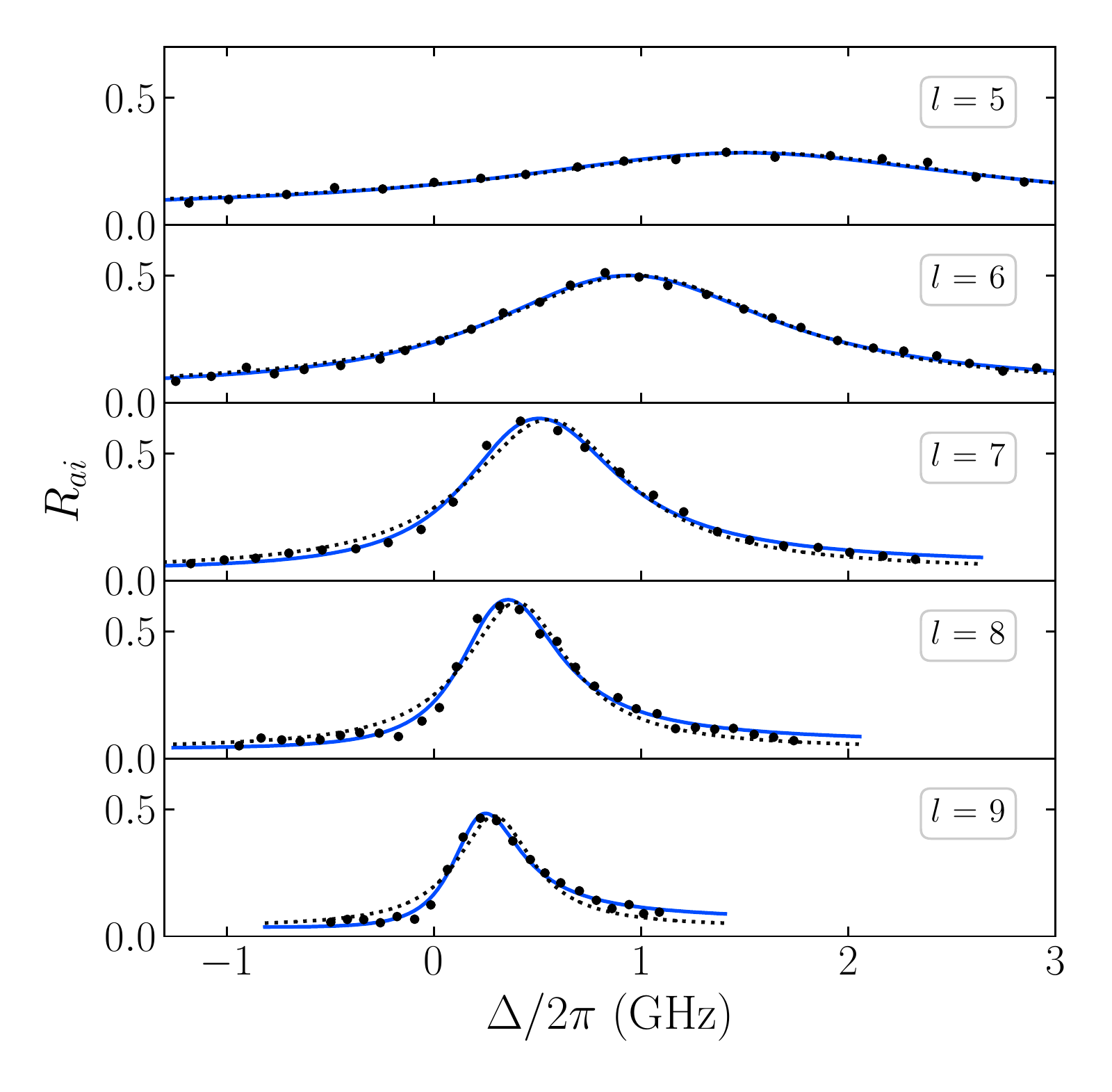}
	\caption{\label{Spec} Autoionization spectra at $n=40$ for various $l$, with a 3 $\mu$s ICE pulse duration. $\Delta$ is the detuning to the $6s_{1/2} \rightarrow 6p_{1/2}$ Yb$^+$ transition. Black circles: experimental points. Dashed grey lines: lorentzian fits. Solid blue lines: fit with asymmetric $g(\Delta)$ function.}
\end{figure}

 On \hyperref[Spec]{figure \ref{Spec}} is plotted the measured ratio of autoionized atoms $R_{ai}=N_{ai}/N_{tot}$ against the detuning to the $6s_{1/2} \rightarrow 6p_{1/2}$ transition of the bare $^{174}$Yb$^+$ ion. Its frequency can be found in \cite{MCL11} and more recently in \cite{ZAL19} which provides the value we used of 811.291 500(40)THz for $^{174}$Yb. As expected, we notice a decrease of the autoionization linewidth and of the overall shift compared to the bare ion core (IC) transition with $l$. This is a clear signature of the decreasing influence of the Yb$^+$ ionic core. Finally, we notice an asymmetry of the spectra for the higher $l$ states. We will first discuss the origin of the asymmetry thanks to an analysis of the spectra.\\

\indent The autoionization profile of doubly excited states is often explained by a quantum interference between two excitation paths: the ionization of the doubly excited state and the direct excitation of the continuum, leading to the so-called Fano profile \cite{FAN61} for the excitation cross-section: 
\begin{equation}
f(\Delta)=\frac{(q\Gamma/2 + \Delta-\Delta_0)^2}{\Gamma^2/4+(\Delta-\Delta_0)^2}
\end{equation} 
where $\Delta_0$ is the pulsation shift compared to the IC resonance, $\Gamma$ is the autoionization rate and $q$ is a symmetry factor, with growing asymmetry for decreasing $q$. As we produce a significant autoionized fraction, the cloud can become depleted which appears as a saturation broadening depending on the pulse duration. The saturation can be modeled with the following fitting function:
\begin{equation}
\label{equation2}
g(\Delta)= \eta \  \bigg( 1-\exp\Big(- \mu \Gamma \tau  \frac{(q \Gamma/2 + \Delta-\Delta_0)^2}{ \Gamma^2/4+(\Delta-\Delta_0)^2}\Big) \bigg)
\end{equation} 
where $\tau$ is the laser pulse length and $\mu$ is a fitted parameter proportional to the photon flux. Finally, we introduce an efficiency parameter $\eta$ which accounts for the geometrical mismatch between the beam and the cloud, and is fixed at the value 0.7. 

However, in the case of the isolated core excitation of Rydberg states, the ICE laser is usually far in the continuum of the IC ground configuration (\textit{i.e.} well above the $I_{6s}$ ionization limit). Photoionisation is then strongly reduced. In that case, the relevant processes are interferences between the excitation of the $6p_{1/2}n l$ state and the neighbouring $6p_{1/2}n'l$ states \cite{TRA82}. For high $l$ states, the autoionization linewidth becomes much narrower than the energy difference to the $6p_{1/2}(n \pm 1) l$ states and the spectrum can be approximated by a simple "saturated" lorentzian \cite{JON88}. 

\begin{figure}[h!]
	\includegraphics[scale=0.45]{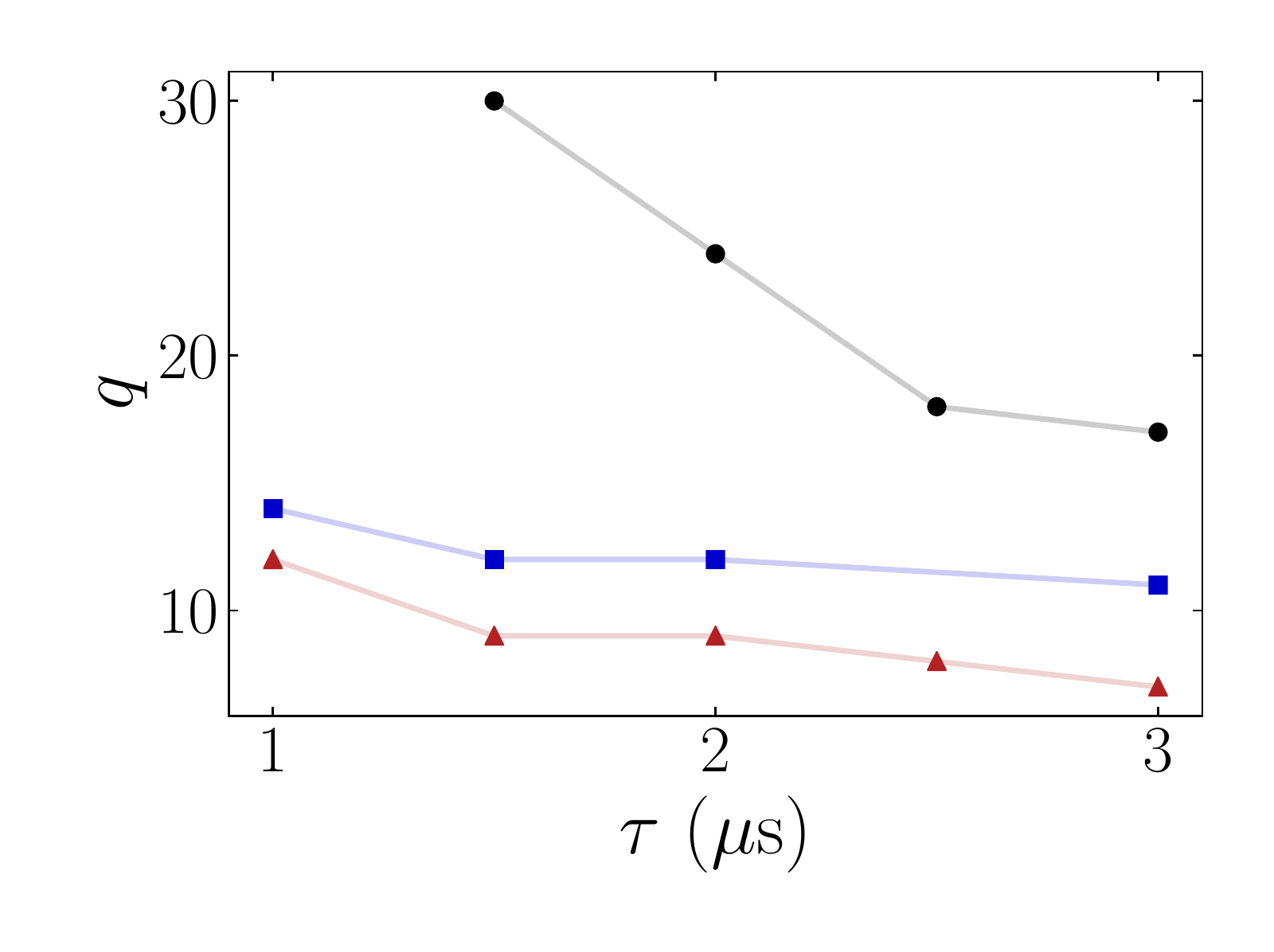}
	\caption{\label{Dep} Evolution of the asymmetry parameter $q$ with ICE pulse duration $\tau$ for $l=7$ (black circle), $l=8$ (blue square), $l=9$ (red triangle).
	} 
\end{figure}

Nevertheless, an asymmetric shape can be seen for the higher $l$ states. In order to understand this feature, one can look at different pulse durations. On \hyperref[Dep]{figure \ref{Dep}}, we can see an increase of the asymmetry with the pulse duration, although it should be independant in the framework of Fano interferences. This result discards a Fano type phenomenon as an explanation for the asymmetry. \\

\indent On the other hand, experimental biases such as local charge effects could explain these asymetries in the spectra. Charges in the cloud, induced by Penning ionizations or autoionization of Rydberg atoms, will mix a fraction of $(l\pm 1)$ character in the $6s_{1/2}nl$ state from Stark effect, thus allowing transitions from $6s_{1/2}nl$ to $6p_{1/2}n (l\pm 1)$ states. The sudden appearence of the charges may also project the atoms into $6s_{1/2}n(l\pm 1)$ states \cite{MIL10}, thus allowing transitions from $6s_{1/2}n(l\pm 1)$ to $6p_{1/2}n (l\pm 1)$ states. Finally, the corresponding Stark effect will shift all these transitions depending on the distance to the charge. The resulting spectrum can then reveal strongly asymmetric profile. Considering our Rydberg atoms density of about $10^7$ cm$^{-3}$, these effects are likely to occur and should increase with the pulse duration due to increasing number of charges. A more systematic study with smaller pulse durations and varying densities might enable to determine exactly which of these effects are actually relevant. Nevertheless, in order to extract the linewidth and shift from the spectra, one cannot use the corresponding multiple sum of Lorentzians as the different peaks are not resolved and it would use too many fitting parameters. Therefore we decided to use the $g$ function anyway, as introduced in \hyperref[equation2]{equation \ref{equation2}}, in the following of this article. This has the advantage of taking into account the depletion effect as well as the asymmetry in a simple function.\\

\begin{figure}[b]
	\includegraphics[scale=0.45]{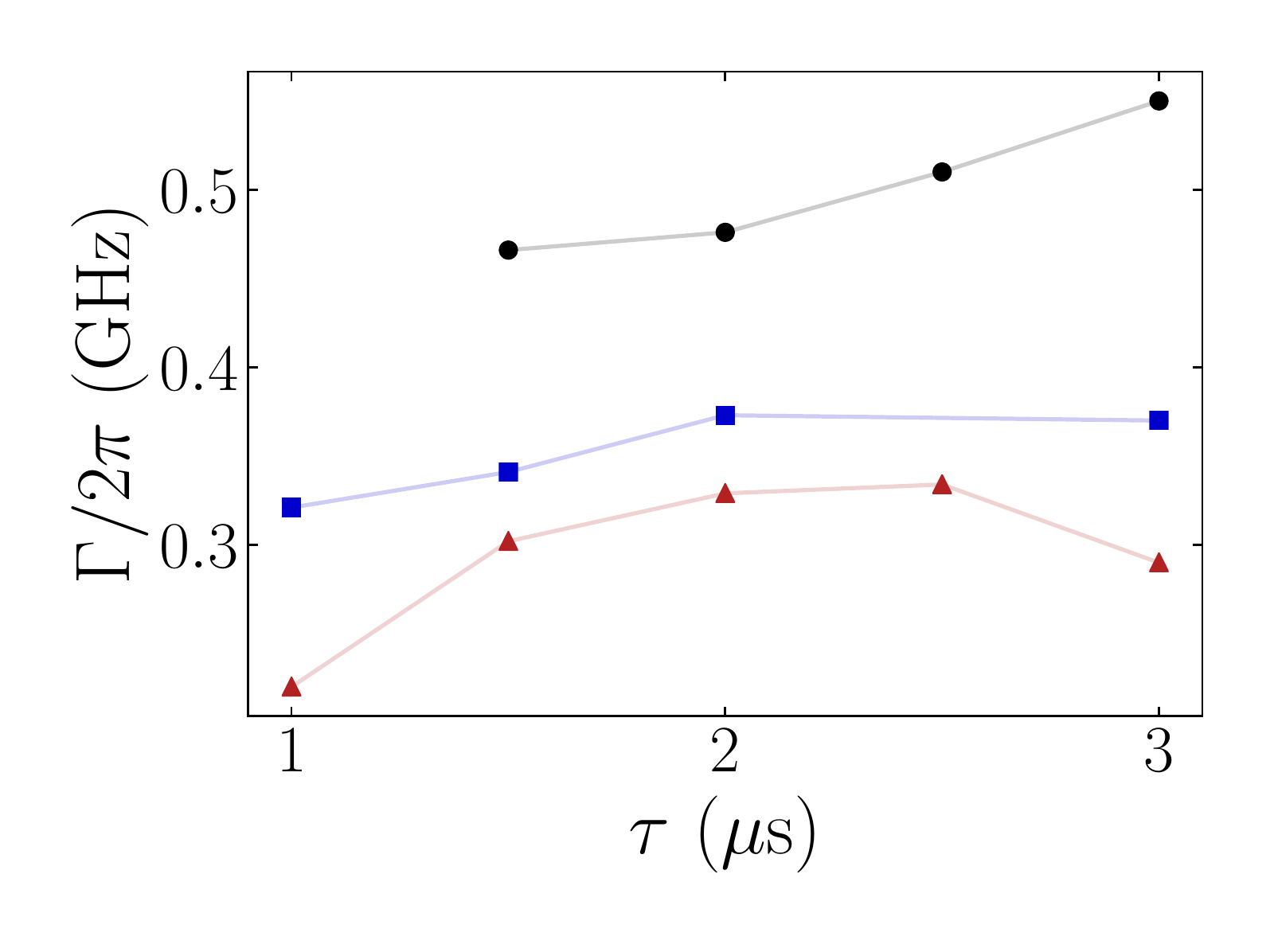}
	\caption{\label{TimeDep} Autoionization linewidth $\Gamma$ as a function of ICE pulse duration $\tau$ for $l=7$ (black circle), $l=8$ (blue square) and $l=9$ (red triangle). Despite taking the depletion into account, we observe a dependance with the pulse duration to be linked with the appearence of asymmetry in the spectra.} 
\end{figure}

Using this fitting function, we extract the autoionization linewidths and the frequency shifts as a function of the pulse duration. The extracted linewidths are displayed on \hyperref[TimeDep]{figure \ref{TimeDep}}. One can see that the autoionization linewidth depends significantly with the pulse durations. This is to be related to the appearence of the unexplained asymmetry and accurate autoionization rates would be obtained by extrapolating the curves towards a zero pulse duration. However the behavior at small pulse duration is unclear and we prefer choosing the linewidth at the smallest measured pulse duration. We then estimate the error by taking the difference with the linewidth at twice this duration. We also add the fitting procedure uncertainty. The obtained experimental linewidths are presented in \hyperref[Linewidth]{figure \ref{Linewidth}} and compared to the theoretical predictions explained in next section. They are also summarized in \hyperref[TableVal]{table \ref{TableVal}}, together with the extracted shifts, which are presented in \hyperref[Energy]{figure \ref{Energy}}. Although the center frequency shifts hardly depend on the pulse duration, we applied the same procedure for their extraction. Another significant error for the center frequency shifts is the absolute uncertainty on the knowledge of the $6s_{1/2} \rightarrow 6p_{1/2}$ frequency of ytterbium ion. This adds a 40 MHz error \cite{ZAL19}.
\begin{table}[h]
	\begin{center}
		\begin{tabular}{|c|c|c|c|cc|}
			\hline
			$\ l \ $& Exp. shift & Th. shift & Exp. width & $ \ \ \ \ \  $ Th.   & width$ \ \ \ \ \  $    \\ 
			&    (MHz)           &   (MHz)    &      (MHz)          & K=$l_2$-1/2, & $l_2$+1/2 \\\hline 
			$\ 5\ $ & 1520(100)  & 2297(600)  & 2570(170)  & 6062        & 2551 \\\hline  
			$\ 6\ $ & 1001(90)  & 1236(230)   & 1175(170) &  2389        & 942 \\\hline  
			$\ 7\ $ & 525(60)  & 650(110)   & 461(70)   &  763        & 292 \\\hline
			$\ 8\ $ & 354(60)  & 358(55)   & 321(60)   &  201         & 76 \\\hline
			$\ 9\ $ & 257(60)  & 207(31)    & 220(90)   &  45         & 17 \\\hline
		\end{tabular} 
		\caption{\label{TableVal} Experimental and theoretical results for the frequency shifts and linewidths for ICE of  Yb on the 6$s_{1/2}$40$l\rightarrow$6$p_{1/2}$40$l$ transition. The theoretical shift is presented for a quadrupolar radial integral $\Bra{6p}r^2\Ket{6p}$=41.7. All frequencies given in MHz. }
	\end{center}	
\end{table}

\section{ \label{Sec4} Theoretical methods}
\label{ThRes}
Thanks to the weak overlap between the inner electron and the Rydberg one, high-$l$ Rydberg states of divalent atoms can be described using a perturbative approach, taking into account two-electron interactions in the frame of the ($j_1 - l_2)K$ coupling scheme where index 1 stands for the inner electron and index 2 stands for the Rydberg electron. We follow previous works performed on barium \cite{JON88,PRU91} and apply it to the case of ytterbium. We consider the following two-valence-electrons-hamiltonian $H = H_1 + H_2 + H_I$ where $H_1$ identifies simply as the independent inner electron hamiltonian of $^{174}$Yb$^+$ ion. $H_2$ corresponds to the Rydberg electron hamiltonian. The third term $H_I ={1}/{r_{12}} - {1}/{r_2}$  accounts for the interactions between inner and outer electrons. We have performed our calculations with the following assumptions:
\begin{itemize}
\item The Rydberg electron radius $r_2$ is much larger than the inner electron radius $r_1$. 
\item High-$l$ Rydberg wave functions are approximated as hydrogenic wavefunctions. 
\end{itemize}

At zero-order, that is to say neglecting the interaction hamiltonian $H_I$, the energy is given by the sum of the two independant electrons energies: $E^{(0)} = E_{n_1l_1j_1} - 1/ {2n_2^2}$. 
The interaction Hamiltonian is then introduced perturbatively. For that matter, we use a Taylor expansion of $1/r_{12}$ in terms of the tensorial operators associated to the spherical harmonics $C^{(q)}(\theta_i,\phi_i)$ \cite{JUD63}.\\

\indent Autoionization rates are calculated using Fermi golden rule. From a $6p_{1/2}nl$ state, only 4 continua contribute to autoionization at first order: $6s_{1/2}\epsilon l\pm 1$ and $5d_{3/2}\epsilon l\pm 1$ where $\epsilon$ corresponds to the energy of the ejected electron. The autoionization rate then reads:
\begin{equation}
\Gamma_{ai} = \sum_{\Bra{\epsilon}} \ |\Bra{\epsilon}\frac{r_1}{r_2^2}C_1^{(1)}C_2^{(1)}\Ket{6p_{1/2}n_2 l_2KM}|^2
\end{equation}
where $\Bra{\epsilon}$ corresponds to the continua previously mentioned. All inner electron dipolar radial integrals were deduced from weighted oscillator strengths in \cite{FAW91} which was chosen for its completeness. The weighted oscillator strength from $6p_{1/2}$ down to $6s_{1/2}$ and $5d_{3/2}$ are 0.465 and 0.126 respectively. We then find the reduced matrix elements $\Bra{6s_{1/2}}|r|\Ket{6p_{1/2}}$ and $\Bra{6p_{1/2}}|r|\Ket{5d_{3/2}}$ to be 2.38 and 3.18 respectively and finally we can deduce the radial integral to be 2.91 and 2.75 in atomic units respectively. These values are similar to the ones from other sources (like \cite{POR12} and references therein). The Rydberg electron integrals are found using the known analytical formula for hydrogen and continuum integrals were calculated with a standard Numerov technique \cite{ZIM79}.\\

\begin{figure}[t]
	\includegraphics[scale=0.47]{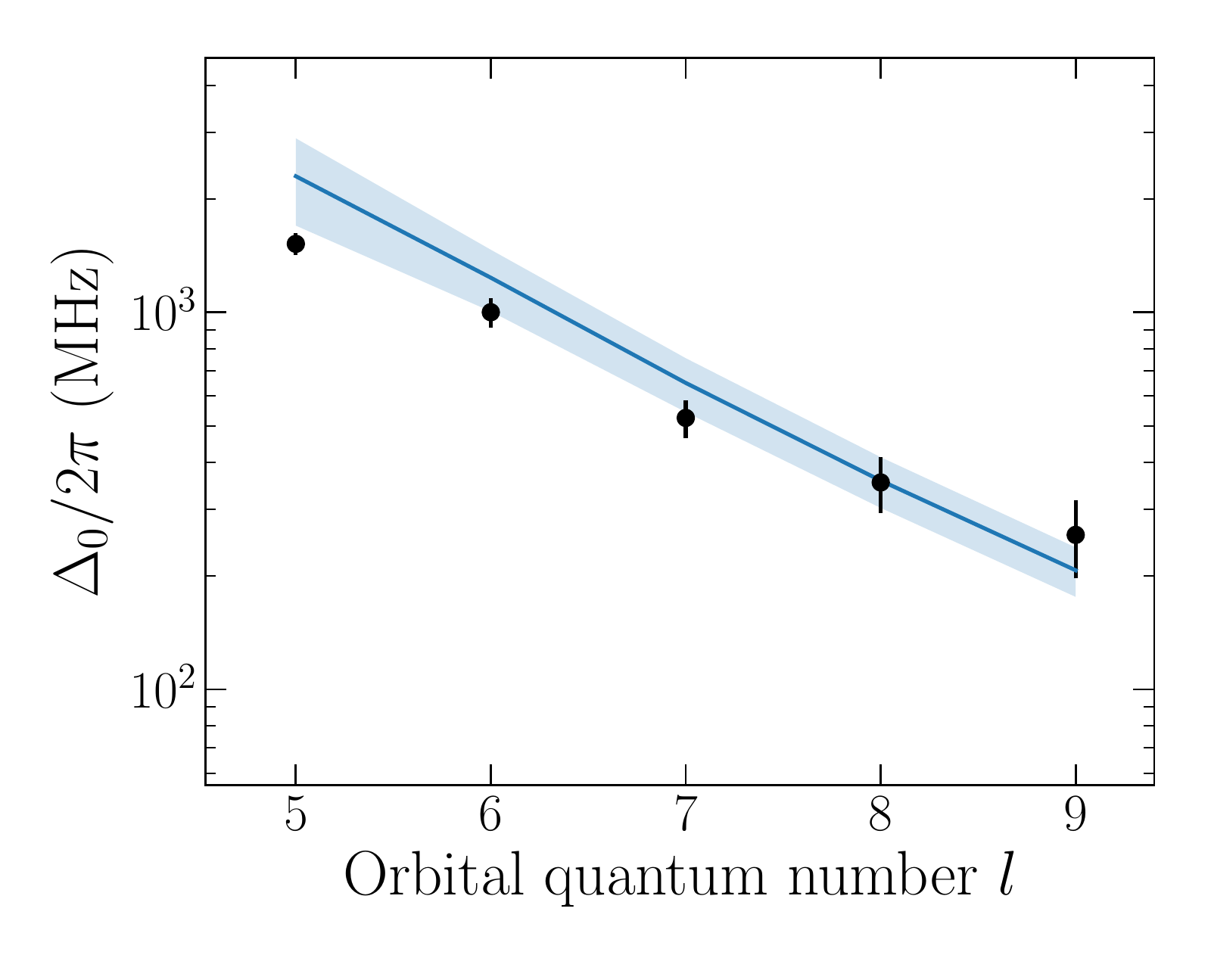}
	\caption{\label{Energy} (Solid lines) Energy shift of the ICE on the $6s_{1/2}40l \rightarrow 6p_{1/2}40l$ transition for the two extremum considered quadratic terms. (Black points) Measured energy shifts. }
\end{figure}

\indent In order to compute the transition energy shift, we compute the shift of both states $6s_{1/2}n_2l_2$ and $6p_{1/2}n_2l_2$ and take the difference. The first order correction to the energy is null and the second order correction is expressed as:
\begin{equation}
E^{(2,q)} = \sum_{q,\Ket{i}} \frac{|\Bra{n_1l_1j_1n_2l_2KM}\frac{r_1^q}{r_2^{q+1}}C_1^{(q)}C_2^{(q)} \Ket{i}|^2}{E_{n_1l_1j_1}- E_i}
\end{equation} 
where $\Ket{i} = \Ket{n_1'l_1'j_1'n_2'l_2' K' M'} $ is every state other than $\Ket{n_1l_1j_1n_2l_2 K M} $ and $E_i = E_{n_1'l_1'j_1'}$ is the zero-th order energy of this state. There, we took into account the closest states that can couple through dipolar ($q=1$) interactions, {\it{i.e.}} all states from \cite{FAW91} for which an oscillator strength is provided (21 and 14 states coupling to $6s_{1/2}$ and $6p_{1/2}$ respectively). They correspond to a complete set of the lowest energy states, including states with an excited $4f$ shell $4f^{13}n_1l_1n_2l_2$, and other higher levels induce a negligible energy shift. Quadrupolar ($q=2$) interactions are typically an order of magnitude lower for $l=5$ and vanish for largest orbital momentums, except for the coupling of $\Ket{6p_{1/2}40 l_2}$ with $\Ket{6p_{3/2}40 l_2'}$ which is particularly close in energy and we add its contribution. The corresponding radial integral could not be found in litterature but realistic boundaries from the reduced matrix elements $\Bra{6p_{1/2}}|r^2|\Ket{6p_{1/2}}=32$ and $\Bra{6p_{3/2}}|r^2|\Ket{6p_{3/2}}=51$ \cite{SAF} can be deduced for the radial integral: $39.2<\Bra{6p}r^2\Ket{6p}<44.2$. Finally we use $41.7\pm5\%$. We also take a $\pm5\%$ uncertainty on the dipolar terms \cite{FAW91} and consider a worst case scenario when summing uncertainties. In \hyperref[Energy]{figure \ref{Energy}}, we show our experimental extracted energy shifts compared to the theoretical shifts within these boundaries. We see a good agreement between theoretical shifts and experimental ones. The residual difference with theory is probably explained by larger uncertainties in the oscillator strengths and missing quadratic terms.

\section{ \label{Sec5} Discussion} 
\label{Dis}
\begin{figure}[t]
	\includegraphics[scale=0.44]{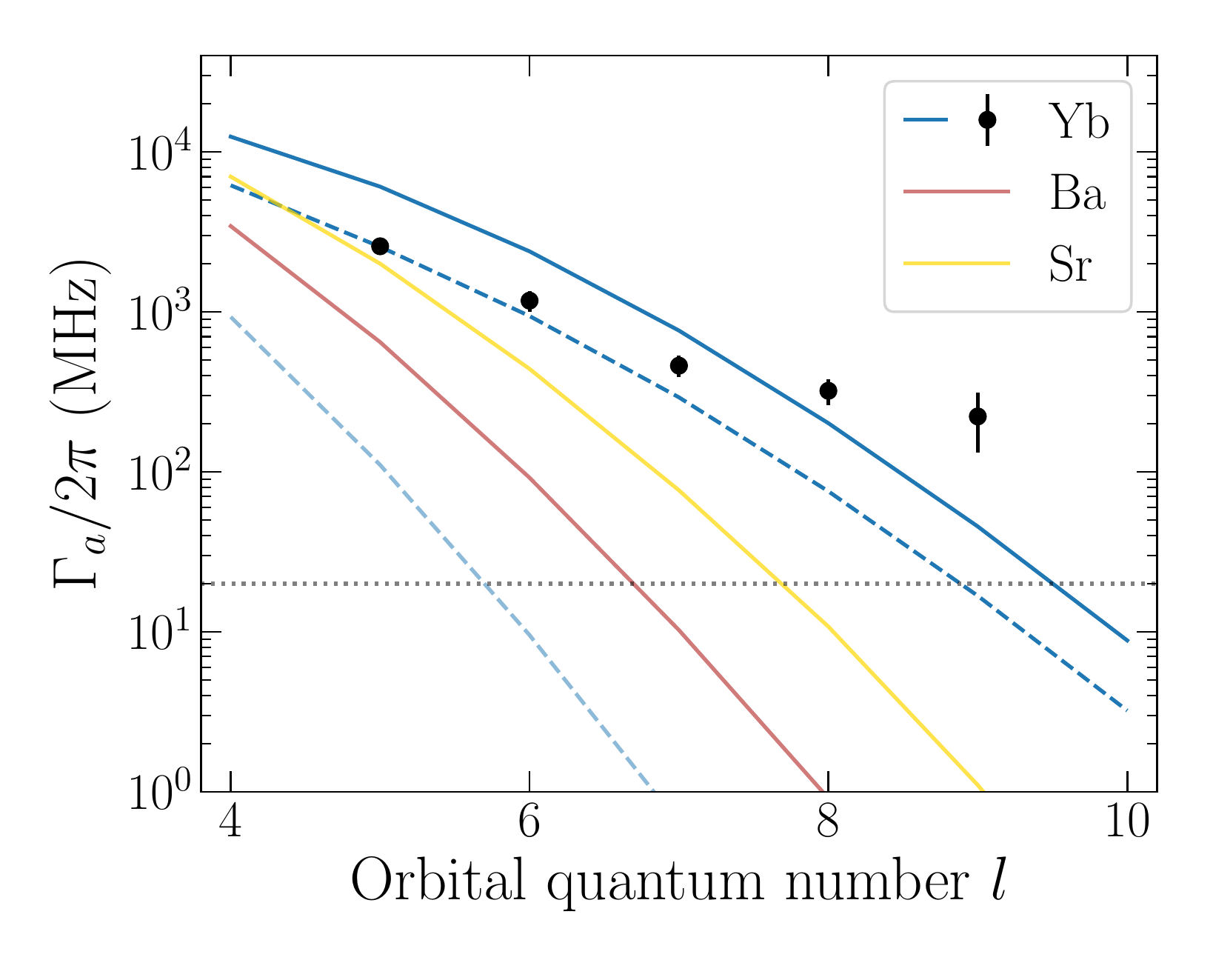}
	\caption{\label{Linewidth} Theoretical linewidths of the ICE spectrum on the $6s_{1/2}40l \rightarrow 6p_{1/2}40l$ of ytterbium for K=$l-1/2$ (blue solid line) and K=$l+1/2$ (dashed blue). We extract the largest ($K=l+1/2$) of the two contributions of $6s_{1/2}\epsilon l$ (light blue dashed lines) and see it is negligible. Comparison with the $6s_{1/2}40l \rightarrow 6p_{1/2}40l$ of barium (red) and the $5s_{1/2}40l \rightarrow 5p_{1/2}40l$ of strontium (yellow), both for $K=l+1/2$. (black dots) Experimental results for ytterbium. (dotted grey line) radiative linewidth for ytterbium of 20 MHz .}
\end{figure}

\indent Due to the relative complexity of the Stark excitation spectrum, we were not able to determine which of the two $K$ state or if a linear combination of these two states was excited. Therefore, we plot in \hyperref[Linewidth]{figure \ref{Linewidth}} the theoretical linewidth of both $K$ states to compare with our results. Extracting the contribution of the sole $\Ket{6s_{1/2}\epsilon_{l'}}$ continua, we clearly see that the coupling to the other two continua $\Ket{5d_{3/2} \epsilon_{l'}}$ dominates the autoionization rate. Indeed, the small energy difference between $5d_{3/2}$ and $6p_{1/2}$ states in Yb$^+$ ionic core implies a larger overlap of the continuum wavefunction with the initial $6p_{1/2}nl$ wavefunction. Comparing with the case of barium and strontium, for which the energy difference is significantly higher, we find that the autoionization rates are indeed much lower. The experimental results compare very well with the theoretical ones especially at low $l$. At larger $l$ the growing mismatch is probably linked to the appearence of the asymmetry in the spectra. Another explanation, not explored here, is the possibility for the Rydberg electron to interact not with just one core electron but with 2 and couple to the ion core state $\Ket{4f^{13}(^2F_{7/2})5d6s(^3D),K=3/2,J=5/2}$ at only around 9 THz below $\Ket{4f^{14}6p_{1/2}}$.\\

\indent We recall our motivation to evaluate the feasibility of \emph{non destructive} imaging of Rydberg atoms by fluorescence of the IC. It is therefore necessary to reduce autoionization below the radiative decay rate. Theoretically, one can see that this condition can be reached for $l \geq 10$ states for $n=40$ states in ytterbium. However, as discussed above, those states are very sensitive to electric field and even a relatively low amount of charge in the cloud can affect the purity of the state. Consequently this imaging technique should probably be restricted to low densities. We also notice that other atoms like barium or strontium are more favorable. Ultimately, it is preferable to choose circular states $n,l=n-1,m=l$ which have the lowest Rydberg electron probability of presence close to the core, and are thus the less autoionizing. Additionnaly, these states have the benefit of being unsensitive to electric fields. Even though the creation of these states demands some efforts \cite{NUS93}, they are promising for an ICE-based imaging technique.

\section{Conclusion} 
In summary, we have investigated both experimentally and theoretically the isolated core excitation of ytterbium atoms in the high orbital $6s40l$ Rydberg states with $l=5-9$. The obtained spectra contained two main features: an unexpected asymmetry for the higher $l$ states, and a relatively low decrease of the linewidth when increasing $l$. The analysis of the experimental spectra shows that asymmetry probably comes from charge induced effects. A theoretical model based on perturbation theory has been used to compute the linewidth and the central energy shifts of the spectra. Due to a strong coupling between the $6p_{1/2}40l$ doubly excited level and the $5d_{3/2}\epsilon_{l'}$ continua, the decrease of the linewidth is found to be relatively slow. This is in good agreement with the experimental measurements. These results lead us to the conclusion that with a good control of the electric fields, a non autoionizing state may be reachable for $l \geq 10$ at $n=40$ and could allow an ICE-based fluorescence imaging. Although, circular states seems ultimately more promising for that application.

\section{Acknowledgements}
The authors would like to thank Tom Gallagher, Laurence Pruvost, Daniel Comparat, Marianna Safronova and Maxence Lepers for fruitfull discussions. This work was supported by the public grant CYRAQS from Labex PALM (ANR-10-LABX0039) and the EU H2020 FET Proactive project RySQ (grant N. 640378).

\bibliographystyle{apsrev4-1}

\bibliography{biblioICE}

\end{document}